\pdfoutput=1 
\documentclass[%
 reprint,
 amsmath,amssymb,
 aps,
]{revtex4-1}

\usepackage{graphicx}
\usepackage{dcolumn}
\usepackage{bm}

\newcommand{\avg}[1]{\left\langle{#1}\right\rangle}

\begin{document}


\title{Reaction networks as systems for resource allocation:\\A variational principle for their non-equilibrium steady states}

\author{Andrea De Martino$^{1,\star}$, 
Daniele De Martino$^{2}$, 
Roberto Mulet$^{3}$, 
Guido Uguzzoni$^{4}$}
\affiliation{%
$^1$ IPCF-CNR, Dipartimento di Fisica, Sapienza Universit\`a di Roma, p.le A. Moro 2, 00185 Roma (Italy)\\
$^2$ Dipartimento di Fisica, Sapienza Universit\`a di Roma, p.le A. Moro 2, 00185 Roma (Italy)\\
$^3$ Group of Complex Systems, Department of Theoretical Physics, Physics Faculty, University of Havana, CP 10400, La Habana (Cuba)\\
$^4$ Dipartimento di Fisica, Universit\`{a} di Parma and INFN Sezione di Parma (Italy)
}%


\begin{abstract}
Within a fully microscopic setting, we derive a variational principle for the non-equilibrium steady states of chemical reaction networks, valid for time-scales over which chemical potentials can be taken to be slowly varying: at stationarity the system minimizes a global function of the reaction fluxes with the form of a Hopfield Hamiltonian with Hebbian couplings, that is explicitly seen to correspond to the rate of decay of entropy production over time. Guided by this analogy, we show that reaction networks can be formally re-cast as systems of interacting reactions that optimize the use of the available compounds by competing for substrates, akin to agents competing for a limited resource in an optimal allocation problem. As an illustration, we analyze the scenario that emerges in two simple cases: that of toy (random) reaction networks and that of a metabolic network model of the human red blood cell.
\end{abstract}

\pacs{Valid PACS appear here}
\maketitle


\section*{Introduction}

The dynamics and thermodynamics of chemical reaction networks is a subject that goes back at least to \cite{prigo,oster3,oster4,oster1,oster2,pere}. Recent years have seen considerable interest in the problem at different levels, from the characterization of their mass-action kinetics \cite{angeli} and of their stochastic thermodynamics based on the chemical master equation \cite{gasp,seif}, to the analysis of their non-equilibrium steady states (NESS) \cite{bqp1,ksch,Zhang:2011fk,Ge:2011uq}. Besides their relevance for fundamental understanding, these approaches provide an important frame for the study of real biochemical systems, like genome-scale reconstructions of cellular metabolic networks \cite{voreed,natr,thiele}. The most basic information about these systems is usually encoded in the matrix of stoichiometric coefficients, representing in essence the (weighted) topology of the couplings between chemical species and reactions. With uncertainties about kinetic parameters and reaction or transport mechanisms often preventing large-scale kinetic approaches (with exceptions like the the metabolism of human erythrocytes \cite{dyna}), the challenge at the simplest level is that of building stoichiometry-based predictive models of metabolic activity at genome scale. Much information on the organization of reaction fluxes in NESS can indeed be obtained from constraint-based models that rely on minimal mass-balance \cite{eip,kau,rev2} or stability \cite{im,martelli} assumptions. Such descriptions revolve around pre-defined (and physically motivated) sets of local constraints, enforcing for instance mass balance at each metabolite node in the network. It would be interesting (and instructive) to {\it derive} the relevant local constraints as the result of a mathematical analysis of the dynamics taking place on the network, both to further clarify the assumptions behind the models and to highlight their limitations. 

This work revisits the joint dynamics of concentrations and reaction fluxes in chemical networks in a statistical mechanics frame. In specific, we obtain a variational principle that relates fluxes (i.e. the average number of microscopic transitions per time per volume for each process) in NESS  to the minima of a global function where stoichiometric coefficients and steady-state concentrations appear as parameters. This function is reminiscent of the Hamiltonian of a Hopfield model with Hebbian couplings \cite{anti}, with stoichiometry playing the role of the patterns. An analysis of its physical meaning explicitly shows that reaction networks dynamically converge towards states where the use of the available compounds is optimized and the rate of decay of entropy production is minimized. The flux organization problem turns out to have remarkable similarities with that of optimal resource allocation by heterogeneous agents, as described e.g. by Minority Games \cite{mg,revmg}. Systems of this type generically undergo a transition from an ergodic phase (the NESS is independent of the initial conditions of the dynamics) to a non-ergodic one when the ratio between the number of reactions and that of chemical species is changed. In our case, the two regimes are described by different sets of local constraints. We shall first explore this scenario in toy ``random'' reaction networks where such a transition can be fully analyzed numerically. Then a simple real system will be considered, namely the reduced metabolic network model of human erythrocytes. Finally, we shall discuss the relevance of these results for the quantitative analysis of cellular metabolism.

\section*{Analysis}

\subsection*{The variational principle}

We consider an open system enclosed in a volume $V$ (a reactor or ``cell'', for brevity), formed by $M$ distinct chemical species (labelled $\mu$) that can be processed by $N$ distinct reactions (labelled $i,j,\ldots$) at fixed temperature $T$, pressure, ionic strength and pH. The reaction stoichiometry is described by the coefficients $\xi_i^\mu$, with the convention that negative (resp. positive) coefficients identify substrates (resp. products) in the `forward' direction of reaction $i$. Individual reaction events occur stochastically with rates (number of events per unit time) proportional to the substrate concentrations, which vary in time. At stationarity and for ideal systems, the Gibbs energy (GE) change per mole associated to each reaction $i$, i.e. \cite{bq}
\begin{equation}\label{uno}
\Delta G_i=\Delta G_{i,0}+RT\sum_{\mu}\xi_i^\mu\log (x^\mu/x_0)
\end{equation}
(where $R$ is the gas constant, $x^\mu$ the intracellular concentration of species $\mu$, $x_0$ a reference concentration and $\Delta G_{i,0}$ the GE change in standard conditions at concentration $x_0$) characterizes its distance from detailed balance. Specifically, the forward ($\phi_{i,+}$) and reverse ($\phi_{i,-}$) fluxes of $i$, i.e. the average number of transitions per time per volume, satisfies the relation $-\beta \Delta G_i=\log(\phi_{i,+}/\phi_{i,-})$ with $1/\beta\equiv RT$ \cite{bqp1}. At equilibrium, $\phi_{i,+}=\phi_{i,-}$ and $\Delta G_i=0$ for each $i$.

We focus on the time evolution of the internal composition of such a chemical reactor. The dynamics of this system is driven essentially by two factors: (a) the fact that molecules can stochastically enter or leave the system,  and (b) the fact that reactions events occur (stochastically) inside the system. Reasonably, then, the NESS will depend (a) on the rate at which molecules can cross the system's boundaries per unit volume (intake or outtake fluxes, denoted by $u^\mu$), and (b) on the rates of the internal reactions (more precisely, on the average net number of microscopic transitions per time per volume, denoted by $\phi_i$). In the following we will show specifically that, given the boundary fluxes $u^\mu$ (characterizing the environment), over time scales for which the chemical potentials can be assumed to vary slowly (so that concentrations can be assumed to be roughly constant) the internal fluxes in a NESS minimize the function
 \begin{equation}
H=\sum_\mu\frac{1}{x^\mu_\infty}\left[\sum_i\xi_i^\mu\phi_i-u^\mu\right]^2\geq 0~~,
\end{equation}
where $x^\mu_\infty$ denotes the concentration of species $\mu$. 

Assume that at time $t=0$ the system is characterized by molecular populations (number of molecules) $n^\mu(0)$.  Consider a time interval of size $\delta t$ and let $\nu_i(t)$ denote the net number of transitions of reaction $i$ that take place between time $t$ and time $t+\delta t$. The latter is a random number governed by a probability law (which shall depend, e.g., on the concentration of substrates) that we leave unspecified for the moment. The corresponding variation of $n^\mu$'s is given by
\begin{gather}
n^\mu(t+\delta t)-n^\mu(t) \equiv \delta n^\mu(t) =\sum_i\xi_i^\mu\nu_i(t)-b^\mu(t)~~,\label{ix}
\end{gather}
where $b^\mu(t)$ stands for the net (random) number of molecules of species $\mu$ taken in (if $b^\mu<0$) or out (if $b^\mu>0$) between time $t$ and time $t+\delta t$. Taking $V$ to be fixed, the time-evolution of concentrations $x^\mu$ is simply 
\begin{gather} \label{x}
x^\mu(t+\delta t)-x^\mu(t) \equiv \delta x^\mu(t) =\frac{\delta n^\mu(t)}{V}~~.
\end{gather} 
We now focus on the quantity
\begin{equation}\label{yps}
y_i(t)=y_{i,0}-\sum_{\mu}\xi_i^\mu\log [x^\mu(t)/x_0]
\end{equation}
(with~ $y_{i,0}=-\beta \Delta G_{i,0}$), whose change between times $t$ and $t+\delta t$ is given by
\begin{equation}
y_i(t+\delta t)-y_i(t) \equiv \delta y_i(t) =-\sum_\mu\xi_i^\mu \log \frac{x^\mu(t+\delta t)}{x^\mu(t)} \label{yy}~~,
\end{equation}
with initial conditions $y_i(0)=y_{i,0}-\sum_{\mu}\xi_i^\mu\log [x^\mu(0)/x_0]$. Equation (\ref{ix}) contains sources of stochasticity in the $\nu_i$'s and the $b^\mu$'s. As a consequence, the ``macroscopic'' variables $x^\mu$ and $y_i$ will fluctuate stochastically as well. Our aim is to characterize the steady state(s) of (\ref{yy}). We make the following simplifying assumptions:
\begin{enumerate}
\item[A1.] Molecular populations are large enough to allow us to treat $x^\mu$ as a continuous variable. This is generically assumed to be the case in real cells, although e.g. in {\it E. coli} the number of copies of certain small molecules can be as low as a few tens (corresponding to a concentration of the order of 10 nM \cite{Milo:2010fk}). The effects induced by molecular noise can be non trivial \cite{seif} and accounting for it might alter the emerging picture \cite{lh}.
\item[A2.] The quantity $\delta x^\mu/x^\mu\simeq \dot{x}^\mu\delta t/x^\mu$ is small (i.e. the chemical potential $g^\mu=g_0^\mu+RT\log x^\mu$, with $g_0^\mu$ the standard chemical potential, is roughly constant), so that $\delta y_i(t)$ can also be taken to be small. 
\end{enumerate}
Under the above assumptions the right-hand side of (\ref{yy}) is easily linearized to yield
\begin{equation}\label{ydot}
y_i(t+\delta t)-y_i(t)\simeq \frac{1}{V}\sum_j \left[-\sum_{\mu}\frac{\xi_i^\mu\xi_j^\mu}{x^\mu(t)}\right]\nu_j(t)+\frac{1}{V}\sum_{\mu}\frac{\xi_i^\mu b^\mu(t)}{x^\mu(t)}~~.
\end{equation}
This equation highlights the way in which concentrations affect the time evolution of the system and, in principle, one would now need to analyze the coupled system formed by (\ref{x}) and (\ref{ydot}). However, for simplicity, we replace $x^\mu(t)$ with some time-independent limit value $x^\mu_\infty$. This approximation can only be justified as long as one considers evolution over time scales shorter than $x^\mu/\dot x^\mu$. If however concentration changes are sufficiently slow (in agreement with homeostasis) it is reasonable to expect that it will hold over time scales much longer than those required to reach a NESS. With this, (\ref{ydot}) takes the form
\begin{equation}\label{ydot2}
y_i(t+\delta t)-y_i(t)\simeq \frac{1}{V}\sum_j J_{ij} \nu_j(t)+\frac{1}{V}\sum_{\mu}\frac{\xi_i^\mu b^\mu(t)}{x^\mu_\infty}~~.
\end{equation}
where the ``couplings'' $J_{ij}$ are defined as
\begin{equation}\label{jay}
J_{ij}=-\sum_{\mu}\frac{\xi_i^\mu\xi_j^\mu}{x^\mu_\infty}
~~~.
\end{equation}
The steady state of (\ref{ydot}) can now be obtained straightforwardly by  dividing both sides by $\delta t$ and averaging over time. This gives
\begin{equation}\label{lang2}
\avg{\frac{\delta y_i}{\delta t}}=\sum_j J_{ij}\avg{\frac{v_j(t)}{V\delta t}}+\sum_\mu\frac{\xi_i^\mu}{x^\mu_\infty}\avg{\frac{b^\mu(t)}{V\delta t}}~~.
\end{equation}
Note that $\avg{v_i(t)/V\delta t}\equiv \phi_i$ is the net flux of reaction $i$ (the average net number of microscopic transitions per time per volume) while $\avg{b^\mu(t)(t)/V\delta t}\equiv u^\mu$ represents the uptake of species $\mu$ (the average number of molecules of species $\mu$ per time per volume entering or leaving the cell). Defining 
\begin{equation}\label{acca}
h_i=\sum_\mu \frac{\xi_i^\mu u^\mu}{x^\mu_\infty}
\end{equation}
we therefore have
\begin{equation}\label{lang}
\avg{\frac{\delta y_i}{\delta t}}=h_i+\sum_j J_{ij}\phi_j\equiv -\frac{1}{2}\frac{\partial H}{\partial\phi_i}~~,
\end{equation}
with
\begin{equation}\label{H}
H=\sum_\mu\frac{1}{x^\mu_\infty}\left[\sum_i\xi_i^\mu\phi_i-u^\mu\right]^2\geq 0~~.
\end{equation}
This implies that, on average, the stochastic dynamics of reactions collectively minimizes the function $H$ so that the NESS of (\ref{lang}) correspond to the solutions of the minimization problem
\begin{equation}\label{min}
\min_{\{\phi_i\}} H~~,
\end{equation}
constrained by the uptake values and by the fact that the $\phi_i$'s are assumed to be bounded by enzyme availability, so that $\phi_i\in[\phi_{i,min},\phi_{i,max}]$. 

Following \cite{mg,revmg}, we can characterize the behavior of fluxes in NESS by noting that the solutions of (\ref{lang}) for $t\to\infty$ (which are expected to be linked to the Gibbs energy change of reaction $i$ in a NESS) are generically of the form $\avg{y_i}\sim \gamma_i t$ with $\gamma_i\equiv h_i+\sum_j J_{ij}\phi_j$, since $\gamma_i$ is constant in NESS. In practice, the stochastic dynamics of $y_i$ for different reactions will be characterized by different values of $\gamma_i$, depending on the asymptotic behavior. If $y_i$ tends to a finite value as $t\to\infty$ then $\gamma_i=0$. Recalling that the Gibbs energy change is related to the forward-to-reverse flux ratio through the detailed balance condition, this describes the case of a reaction whose microscopic transitions occur bidirectionally even for $t\to\infty$, i.e. such that the forward and reverse fluxes are both non-zero in the steady state. Specifically, its flux is determined by the condition $h_i=-\sum_j J_{ij}\phi_j$. When $\gamma_i\neq 0$, instead, $y_i$ increases or decreases linearly in time (after a transient) and diverges for $t\to\infty$ so that, in the corresponding NESS, reaction $i$ will be characterized by unidirectional microscopic transitions for $t\to\infty$.  Note that if $\gamma_i=0$ then $H$ is indeed minimized by solving $\partial H/\partial\phi_i\equiv -2\gamma_i=0$, whereas for $\gamma_i\neq 0$ the minimum of $H$ is achieved by taking $\phi_i$ as large (if $\gamma_i>0$) or as small (if $\gamma_i<0$) as possible, i.e. $\phi_i=\phi_{i,max}$ (resp. $\phi_i=\phi_{i,min}$) if $\gamma_i>0$ (resp. $\gamma_i<0$).

\subsection*{Physical meaning of $H$}

For a start, notice that the change in $y_i$ is expressed in equation (\ref{ydot2}) as the sum of two terms. The first accounts for the coupling of $i$ to other reactions in the network via shared compounds (the coupling coefficient $J_{ij}$ is non-zero only if $i$ and $j$ have a metabolite with finite concentration in common). Consider a species $\mu$ and two reactions $i$ and $j$ such that $\xi_j^\mu>0$ ($j$ produces $\mu$) and $\xi_i^\mu<0$ ($\mu$ is consumed by $i$). Then $J_{ij}>0$ and a positive net advancement of reaction $j$ will contribute to the increase of $y_i$, see (\ref{ydot2}). Now looking at (\ref{yy}) it is reasonable to expect that the probability of observing a forward transition for reaction $i$ between time $t$ and time $t+\delta t$ will be larger the larger (and positive) is $y_i(t)$ (in agreement with the fluctuation theorem \cite{gasp}; we shall see an explicit example of this later on). Therefore in this case a positive net advancement of reaction $j$ will ultimately increase the probability of a concomitant advancement of reaction $i$. In other words, this situation favors the emergence of a positive correlation between reactions $i$ and $j$. Likewise, if both $i$ and $j$ are either producing ($\xi_i^\mu>0$, $\xi_j^\mu>0$) or consuming ($\xi_i^\mu<0$, $\xi_j^\mu<0$) species $\mu$, their coupling will tend to anti-correlate $\nu_i$ and $\nu_j$. In this case the dynamics discourages over-production or over-consumption of a chemical species. A similar picture holds for the coefficients $h_i$, which are related to the presence of sources (like nutrients) and sinks (e.g. outtakes) in the network. A non-zero $h_i$ acts as a force of magnitude proportional to $|u^\mu|$ that tends to polarize in a particular direction a reaction $i$ that is stoichiometrically connected to a compound $\mu$ with $u^\mu\neq 0$. The favoured direction depends on whether $\mu$ is a source or a sink and on the sign of $\xi_i^\mu$. This effect can propagate to other nodes connected to $i$  if $|u^\mu|$ is sufficiently large. The role of the concentrations $x^\mu_\infty$ appearing in $J_{ij}$ and $h_i$ is in essence that of modulating the strength of the couplings and of the forcing fields with the availability of the intermediate metabolites. Indeed, $J_{ij}$'s get stronger when the intermediate compounds are present in smaller amounts, stressing the emergence of positive correlations between processes (if $J_{ij}>0$) or the limits imposed by competition for a limited resource (if $J_{ij}<0$). When the concentration of the intermediate is large, instead, the coupling gets weaker and $i$ and $j$ may become effectively independent. The impact of $h_i$ is understood along similar lines. In this way, the original bipartite network of reactions and metabolites can be re-cast as a system of interacting reactions with Hebbian couplings, as shown in Fig. \ref{uno}. 
\begin{figure}[!]
\begin{center}
\includegraphics[width=8cm]{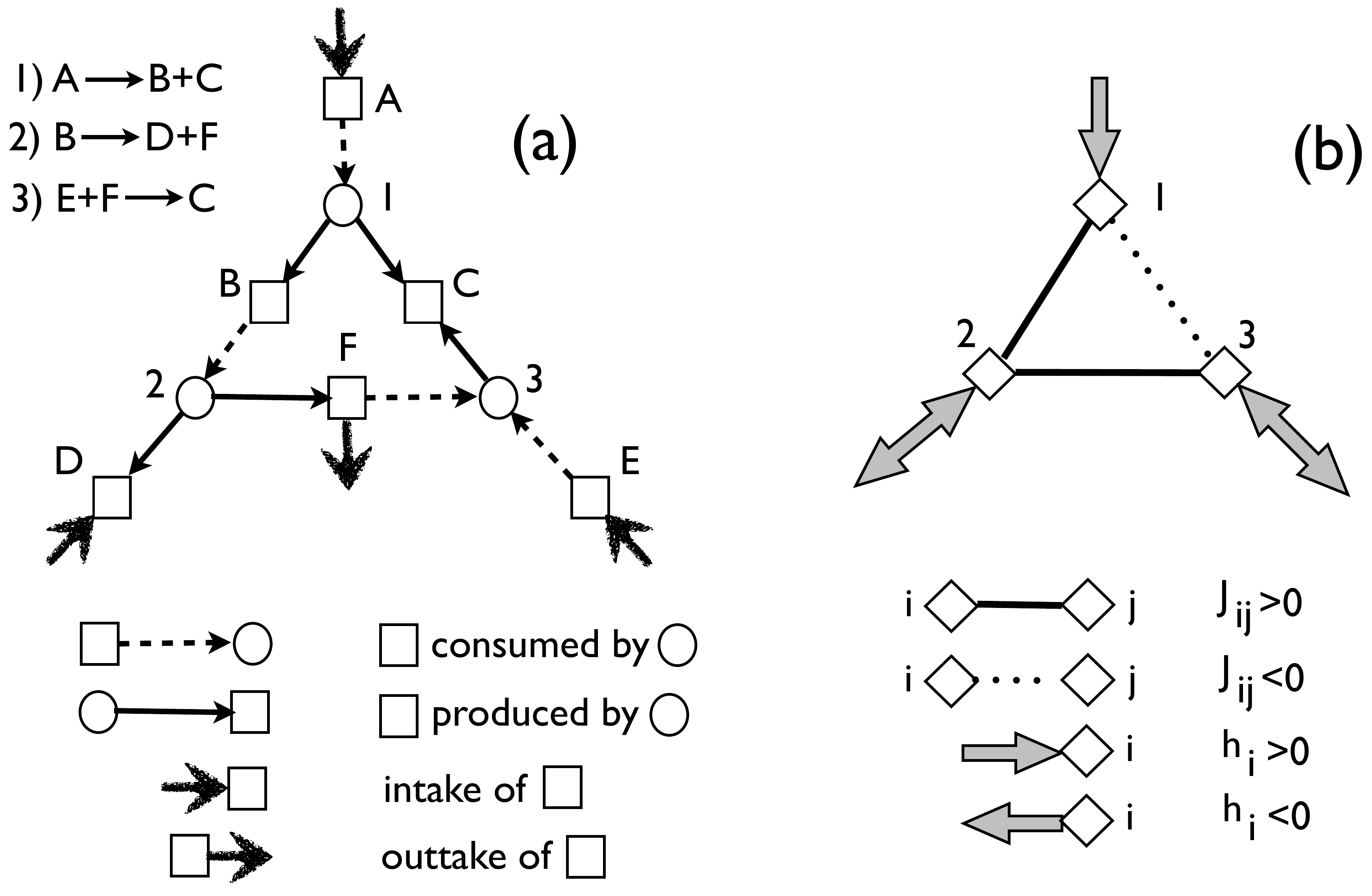}
\caption{\label{uno} 
{\bf (a)} Toy reaction network with reactions represented as circles and chemical species as squares. Continuous, dashed, incoming drawn and outgoing drawn arrows denote stoichiometric coefficients and uptakes, respectively $\xi_i^\mu>0$, $\xi_i^\mu<0$, $u^\mu<0$ and $u^\mu>0$. {\bf (b)} Reduced reaction network with couplings and ``fields'' given by (\ref{jay}). Continuous, dotted, incoming grey and outgoing grey arrows denote respectively $J_{ij}>0$, $J_{ij}<0$, $h_i>0$ and $h_i<0$. For instance, $J_{13}=-\xi_1^C\xi_3^C/x^C<0$, $h_1=\xi_1^A u^A/x^A>0$. Grey arrows are double-headed when the sign of $h$ depends on the precise values of stoichiometric coefficients and uptake fluxes. For instance, the value of $h_2=\xi_2^D u^D/x^D+\xi_2^F u^F/x^F$ depends on the choice of the  $\xi$'s and $u$'s, since the first term in the sum is negative while the second is positive.}
\end{center}
\end{figure}
This is strongly reminiscent of Hopfield models of neural networks. In such a scenario, finding the steady state(s) of (\ref{ydot}) is equivalent to finding the ground states of a system of reactions interacting with ``energy'' $H$.

From a physical standpoint, $H$ quantifies the resource mis-usage by the network so that, by minimizing $H$, the system strives to reach states in which compounds are used as optimally as possible, given the initial conditions $y_i(0)$ (that also account for the standard GEs and hence, to some degree, for the {\it a priori} reversibility), the stoichiometry, the available nutrients, the production goals, etc. Whether for a {\it given network} the minimum of $H$ is zero or not then depends on several factors, including the bounds on fluxes and the specific form of the uptakes. Note that $\min H=0$ would imply that at stationarity 
\begin{equation}\label{mb}
\sum_i\xi_i^\mu\phi_i=u^\mu~~~~~\forall\mu~~,
\end{equation}
i.e. that in the steady state a Kirchhoff-law type of scenario holds in which strict mass-balance conditions are satisfied for each chemical species. The network in this case organizes the fluxes so that consumption and production exactly match for each species and meet the nutrient availability and outtake requirements described by the vector $\{u^\mu\}$. On the other hand, the physically relevant states with $\min H\geq 0$ can more generally be thought to have
\begin{equation}\label{vn}
\sum_i\xi_i^\mu\phi_i\geq u^\mu~~~~~\forall\mu~~,
\end{equation}
otherwise the system could e.g. consume a nutrient in excess of its availability.  Both sets of conditions have been employed for the modeling of cellular metabolic networks. In particular equations of the type of (\ref{mb}) are the basis of the highly successful flux-balance-analysis (FBA) \cite{kau,rev2}, where biological functionality is included as an additional {\it ad hoc} constraint usually represented by the maximization of a specific score function (e.g. biomass production for {\it E. coli} in an optimal environment). The conditions (\ref{vn}) are instead reminiscent of Von Neumann's model of reaction networks \cite{em}, where self-consistent flux states with a net positive production of intracellular metabolites can be allowed. This type of approach can be helpful in analyzing the metabolic capabilities of an organism and, if statistically robust production profiles emerge, in inferring (rather than postulating) cellular objective functions, with the idea that chemical species that are globally produced (e.g. amino acids) are employed in macromolecular processes (like protein formation) that are not encoded by the reaction stoichiometry. We remark however that within the above setting finding the NESS of the system means minimizing $H$, which obviously is a priori different from solving (\ref{mb}) or (\ref{vn}).

A broader physical insight can be obtained by studying how $H$ relates to standard quantities used to characterize non-equilibrium behavior in reaction networks, such as the entropy production. Indeed the entropy production per volume $\dot{S}_{int}$ of the system enclosed in the ``cell'' of volume $V$ is given by 
\begin{equation}
T\dot{S}_{int}=-\sum_\mu \dot{x^\mu} g^\mu~~,
\end{equation}
where $x^\mu$ is the concentration of species $\mu$ and $g^\mu=g_0^\mu+RT\log x^\mu$ is its chemical potential. Taking time derivatives keeping in mind that at stationarity $\dot{x^\mu}$ is constant, one easily finds that $H\simeq \sum_\mu (\dot{x}^\mu)^2/x^\mu=-\ddot{S}_{int}/R$. Now since $H\geq 0$, $\ddot{S}_{int}\leq 0$. In addition, $H$ is constant in NESS. It follows that for time scales shorter than $x^\mu/\dot{x}^\mu$ (i.e. for time scales over which chemical potentials are roughly constant) one has 
\begin{equation}
\dot{S}_{int}\simeq \dot{S}_{int}(0)-RHt~~.
\end{equation}
$H$ is thus seen to play the role of the rate at which the entropy production changes in a NESS. If $\min H=0$ (i.e. if the NESS is flux-balanced), then  $\dot{S}_{int}$ must vanish as well (since $T \dot{S}_{int}=-\sum_\mu g^\mu(\sum_i\xi_i^\mu\phi_i-u^\mu)$), leading to a state with constant entropy. If $\min H>0$, instead, the entropy production is non-zero and decreases at the smallest allowed rate. Correspondingly, the entropy ``slows down'' quadratically over the time scales for which the theory holds. (Note that for sufficiently long times the entropy production can become negative if $H>0$, implying that this limit may be unphysical.) In summary, the NESS that can be obtained are either characterized by $H=0$, zero entropy production and hence constant entropy, or by $H>0$, positive entropy production decreasing in time as slowly as possible and entropy increasing in time accordingly. 

It is noteworthy that this scenario essentially characterizes the Lyapunov condition described in \cite{Kondepudi:1998fk} for the stability of stationary states (Chapter 18). The Lyapunov function in our case can be computed explicitly, takes the form (\ref{H}), its minimization bears a further physical meaning in terms of optimal resource allocation, and provides an equivalent description of the reactor as a system of processes interacting via Hebbian couplings. Indeed the quantity $H$ appears in the Gibbs theory of thermodynamic stability for ideal systems. Let us consider a system at equilibrium with vanishing net fluxes and evaluate the effect of a perturbation that drives the system away from  equilibrium. Such a perturbation corresponds to a (small) change in the turnover $\delta z_i$ of each reaction $i$, which can be achieved e.g. by forcing a (small) non-zero flux $\phi_i$ through each reaction $i$ for a time $\delta t$, so that $\delta z_i\propto \phi_i\delta t$. The free energy per volume associated to the perturbed state is easily found to be given by
\begin{equation}
G\simeq G_{eq}+ RTH\delta t^2 > G_{eq}~~,
\end{equation}
in agreement with the second law of thermodynamics. After turning off the perturbation, the system will relax back to equilibrium by minimizing $G$ (i.e. $H$).

\section*{Results}

\subsection*{A toy model: random reaction networks}

A simple algebraic argument allows to understand that, generically, a qualitative change in the solutions of (\ref{min}) is expected to take place as one varies the ratio between the network parameters $N$ and $M$. Let $N_{{\rm rev}}$ denote the number of reactions that remain asymptotically bidirectional, i.e. such that $\gamma_i=0$. The conditions $\sum_j J_{ij}\phi_j=-h_i$ that they must satisfy form a set of $N_{{\rm rev}}$ equations, $M$ of which at most are independent (strictly speaking, the rank of the matrix $\{J_{ij}\}$ equals that of the stoichiometric matrix $\{\xi_i^\mu\}$, see (\ref{jay}); it is however always possible to eliminate dependencies in the latter so as to attain full rows rank). Neglecting the fact that variables are bounded, we can say that when the number of variables ($N_{{\rm rev}}$) exceeds that of equations the system is underconstrained and multiple solutions occur. So  multiple steady states exist when $M<N_{{\rm rev}}$. Note that $N_{{\rm rev}}$ is determined {\it dynamically} by (\ref{ydot}) with initial conditions $y_i(0)$. This suggests that different choices of $y_i(0)$ will lead to different steady states, i.e. that when $M<N_{{\rm rev}}$ ergodicity (i.e. independence of the steady state on initial conditions) won't hold. A transition is thus expected to occur when $N_{{\rm rev}}=M$.

Such a transition can be fully investigated within the following, simple model: the stoichiometric coefficients $\{\xi_i^\mu\}$ are independent, identically distributed random numbers (such that each reaction uses and produces a finite fraction of the possible compounds), and the dynamics of the network advances in discrete time steps of size $\delta t$. In specific, at each time $t$ the $\nu_i$'s take on values stochastically in $\{-1,1\}$ with
\begin{equation}\label{db}
\frac{\text{Prob}\{\nu_i(t)=1\}}{\text{Prob}\{\nu_i(t)=-1\}}=e^{y_i(t)}=e^{y_{i,0}}\prod_\mu x^\mu(t)^{-\xi_i^\mu}~~,
\end{equation}
i.e. with $\text{Prob}\{\nu_i(t)=\pm 1\}\propto \exp[\pm y_i(t)/2]$. Note that the above probability ratio is proportional to the ratio between the concentration of substrates and that of products of the reaction. Clearly, (\ref{db}) does not allow to capture the temporal structure induced by the Arrhenius law, because by forcing each reaction to take place at every time step it neglects the fact that activation energies (and hence characteristic timescales) can differ significantly across reactions. It is however reasonable to think that the steady state will be unaffected by these transients. On the other hand (\ref{db}) makes the theory considerably easier from a mathematical viewpoint and the final result for the steady state (which is the focus of the present work) more transparent. Setting $V=\delta t=1$ for simplicity, so that $-1\leq\phi_i\leq 1$ for each $i$, one can follow \cite{ctl} to derive the continuous-time limit of (\ref{ydot}) for large $N$ and $M$. The result is the Langevin process
\begin{equation}\label{langu}
\dot{y_i}=h_i+\sum_j J_{ij} \tanh(y_j/2)+\eta_i~~,
\end{equation}
where $\eta_i$ is a Gaussian noise with zero mean. Clearly, time-averaging leads back to (\ref{lang}) with $\phi_i=\avg{\tanh(y_i/2)}$. Recalling that $-\beta \Delta G_i=\log(\phi_{i,+}/\phi_{i,-})$ and noting that this implies $\phi_i\equiv \phi_{i,+}-\phi_{i,-}=(\phi_{i,+}+\phi_{i,-})\tanh (-\beta\Delta G_i/2)$ where, by (\ref{db}), $\phi_{i,+}+\phi_{i,-}=1$, this in turn suggests the relation
\begin{eqnarray}
\Delta G_i & = & -\frac{2}{\beta}\,\text{arctanh}\avg{\tanh(y_i/2)}\nonumber\\
& \equiv &  -\frac{1}{\beta}\log\frac{1+\avg{\tanh(y_i/2)}}{1-\avg{\tanh(y_i/2)}}~~,
\end{eqnarray}
which explicitly links the thermodynamic driving force of a reaction to the (stochastic) dynamics of the quantity $y_i$. The fact that the final result differs from the na\"ive intuition $-\beta \Delta G_i=\avg{y_i}$ is a direct consequence of the stochastic fluctuations encoded in (\ref{db}). Note that the dynamics (\ref{db}) thus converges to NESS (minima of $H$) that are thermodynamically feasible, in agreement with the second law of thermodynamics. We can study the linear stability of (\ref{langu}) by setting
\begin{equation}
y_i(t)=2\,\text{arctanh}(\phi_i^\star)+\lambda_i(t)
\end{equation}
where $\{\phi_i^\star\}$ is a NESS and $\lambda_i(t)$ is a zero-average noise representing (small) perturbations to the trajectory. Reactions with $\phi_i^\star=\pm 1$, for which $y_i$ diverges, will be insensitive to $\lambda_i(t)$. Hence it suffices to focus our attention on the response of reactions for which $(\phi_i^\star)^2<1$. To first order in $\lambda_i$, fluctuations are easily found to obey the condition
\begin{gather}
\dot{\lambda_i}=-\frac{1}{2}\sum_j S_{ij}\lambda_j\\
S_{ij}=-J_{ij}[1-\tanh^2(y^\star_j/2)]~~,
\end{gather}
where $y^\star_i=2\,\text{arctanh}(\phi_i^\star)$. Now if all eigenvalues of the matrix $S_{ij}$ are positive the dynamical system will be linearly stable as small perturbations occurred along the trajectories will die out in time. The term $1-\tanh^2(\overline{y}_j/2)$ is clearly positive, hence it suffices to check that the smallest eigenvalue of the matrix $-J_{ij}=\sum_\mu\xi_i^\mu\xi_j^\mu/x^\mu_\infty$ is positive. Assuming that $\xi_i^\mu$ are independently and identically distributed, the spectrum of $J_{ij}$ can be computed, in the limit $N\to\infty$ with $n=N/M$ finite, using the results of \cite{sm}. For the smallest eigenvalue one finds $\lambda_{min}=a^2\left[1-\sqrt{n_{{\rm rev}}}\right]$, where $n_{{\rm rev}}=N_{{\rm rev}}/M$ and $a^2$ is a constant. Hence stability (and ergodicity) requires $n_{{\rm rev}}<1$. When $n_{{\rm rev}}>1$, instead, the dynamics will be sensitive to small perturbations and, reasonably, its steady state will be selected by the initial conditions $y_i(0)$. The marginal stability condition $n_{{\rm rev}}=1$ coincides with the rough algebraic estimate of the transition point made above, i.e. $N_{{\rm rev}}=M$.

\begin{figure}[!]
\begin{center}
\includegraphics[width=8cm]{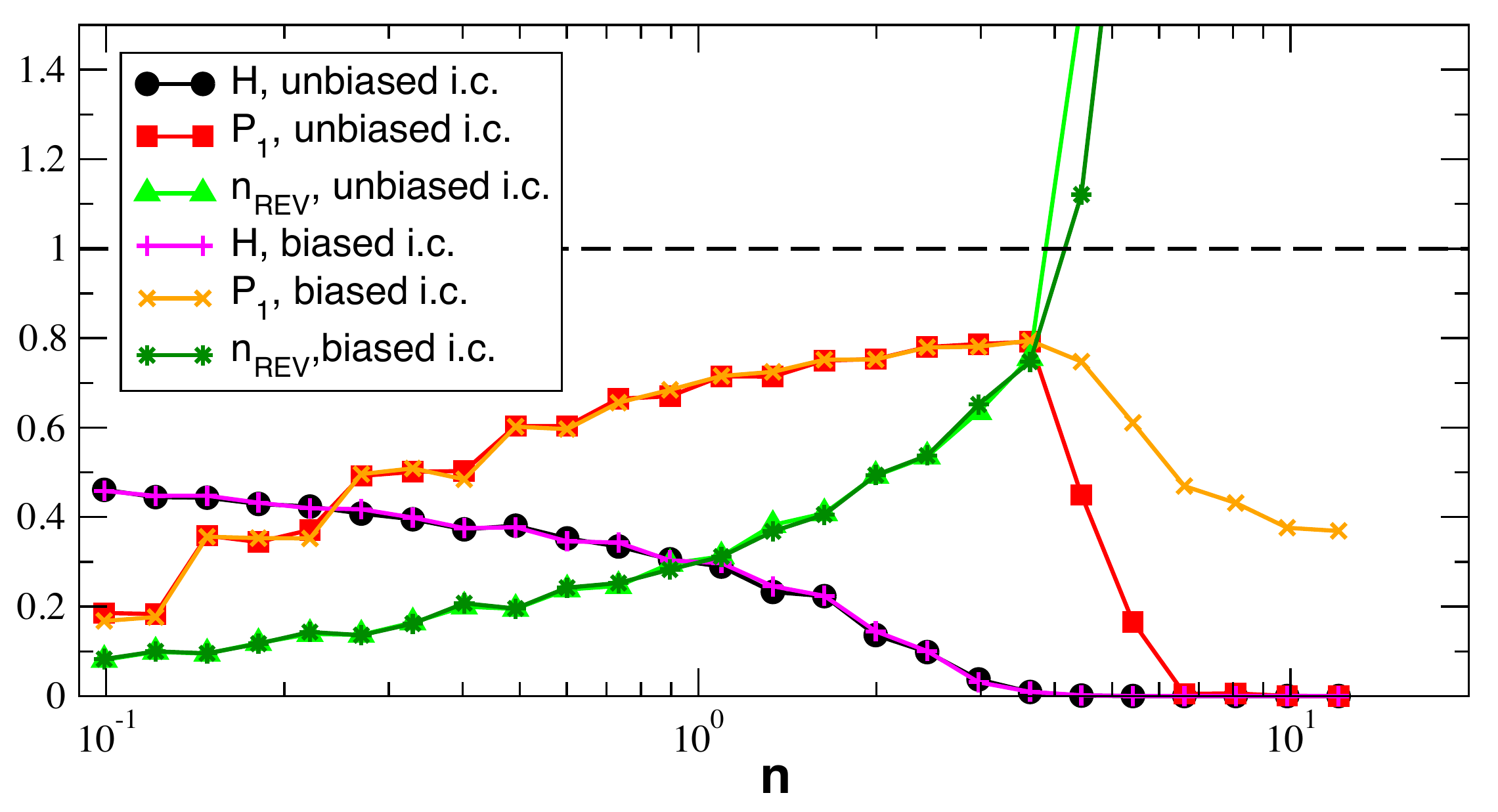}
\caption{\label{due}
{\bf Figure 2.} Average stationary values of $H$, fraction of asymptotically unidirectional reactions ($P_1$) and relative number of asymptotically bidirectional reactions $n_{{\rm rev}}=N_{{\rm rev}}/M$ versus $n=N/M$ obtained from (\ref{dynam}) (with no prior assumption on reaction reversibility for an ensemble of random reaction networks with $NM=10^4$ constructed as described in the text. Averages are taken over 200 realizations for each value of $n$. Unbiased i.c. (initial conditions) refers to steady states of (\ref{dynam}) for $y_i(0)=0$; biased i.c. instead correspond to $y_i(0)=5N$ for each $i$. Note the two phases with $H>0$ ($n<n_c)$ and $H=0$ ($n>n_c$) as predicted. The critical point $n_c$ coincides within numerical error with the point where $N_{{\rm rev}}=1$. Finally, the phase with $H=0$ is non-ergodic: different initial conditions lead to different NESS, characterized by different values of $P_1$.}
\end{center}
\end{figure}

To illustrate the scenario underpinned by the above theory, we have simulated the dynamics defined by
\begin{equation}\label{dynam}
y_i(t+1)-y_i(t)= \sum_j J_{ij}\nu_j(t)+h_i~~~,
\end{equation}
for an ensemble of artificial reactors formed by $M$ species and $N$ reactions in order to analyze the dependence of its steady state(s) on $n=N/M$. Stoichiometric coefficients were chosen randomly, so that $\xi_i^\mu=0$ with probability $p$ and $\xi_i^\mu=\pm 1$ with probabilities $(1-p)/2$, independently on $i$ and $\mu$. Similarly, for the boundary metabolites we took $u^\mu=0$ with probability $q$ and $u^\mu=\pm 1$ with probability $(1-q)/2$, independently on $\mu$. For sakes of simplicity, we set $x^\mu_\infty=1$ for all $\mu$ (it is clear that the choice of $x^\mu_\infty$ does not affect the transition point where $H$ vanishes; it only changes the value of $\min H$ in the ergodic phase). No prior assumption on reversibility was made, i.e. all microscopic transitions can initially occur in both directions for each process. The coefficients $J_{ij}$ and $h_i$ are defined as in (\ref{jay}) and (\ref{acca}) except for the fact that $J_{ij}$ is re-scaled by $\sqrt{N}$ to ensure that the system is well-behaved when $N\gg 1$, while $\nu_i(t)$ evolves according to (\ref{db}). For consistency, $H$ for this case is defined as
\begin{equation}
H=\frac{1}{M}\sum_\mu\frac{1}{x^\mu_\infty}\left[\frac{1}{\sqrt{N}}\sum_i\xi_i^\mu\phi_i-u^\mu\right]^2~~.
\end{equation}
Results for $p=q=1/2$ are shown in Figure ~\ref{due}. 
One sees that initializing the dynamics from ``equilibrium'' conditions $y_i(0)=0$ for each $i$ one ends up in stationary states with $H>0$ for $n<n_c\simeq 3$ whereas $H=0$ for $n>n_c$.  Similarly, $n_{{\rm rev}}<1$ for $n<n_c$ while $n_{{\rm rev}}>1$ for $n>n_c$, so that the critical point is indeed marked by the condition $n_{{\rm rev}}=1$. When the dynamics starts from $y_i(0)=5N$ for each $i$, instead, the system ends up in a different steady state for $n>n_c$, as signaled by the different value of the fraction of asymptotically unidirectional processes at stationarity, $P_1=(N-N_{{\rm rev}})/N$. In the phase characterized by $H>0$, the steady state is unchanged. These results fully confirm the theoretical predictions derived above. We also note that (data not shown), if $u^\mu=0$ for each $\mu$, i.e. if there is no exchange with the surroundings, the corresponding networks converge to a steady state in which $\phi_i=0$ for each $i$,  $H=0$ and $P_1=0$ for each $n$, i.e. to chemical equilibrium. In other words, expectedly, boundary fluxes induce NESS.

\subsection*{Red cell metabolism}

We now turn to a somewhat more realistic case in which nevertheless a full analytic study of the scenario underlying the minimization of $H$ is possible, namely the standard reduced model of the hRBC metabolism, which includes the glycolytic and the pentose phosphate pathway only (21 reactions among 30 metabolites plus two ionic pumps, namely ATPase and NADPHase; see Tables I and II for the numerical details and the network structure used here). They operate by consuming glucose (GLC) in order to, respectively, maintain the osmotic balance through the sodium-potassium ionic pump (ATPase), and reduce the amount of free radicals through glutathione reductase (NADPHase). Moreover, at a simplified level, one can think that the final products of these pathways are, respectively, lactate (LAC) plus an exchange of K$^+$ and Na$^+$ ions, and CO$_2$. The partitioning of GLC between the two pathways depends on the level of oxidative stress faced by the cell. Experimental estimates based on enzyme activities range from 70\% or more in favor of glycolysis in unstressed conditions to 70\% or more in favor of the pentose-phosphate pathway under oxidative stress \cite{Milo:2010fk}. We want to use the theory described here to estimate the range of variability of the fraction of glucose consumed by each pathway as a function of the oxygen concentration in the environment. 

We start by assuming (reasonably) that the steady state of the hRBC metabolism is compatible with $\min H=0$ (i.e. flux balance). A straightforward analysis of the emerging equations reveal that only three of the 23 reactions are linearly independent: we choose the glucose uptake $u^{{\rm GLC}}$, the flux through the Rapoport-Leubering shunt (or through the enzyme 2,3-DPG mutase, DPGM, a key step that regulates the haemoglobin's affinity with oxygen) $\phi_{\text{DPGM}}\equiv \phi_{\text{RLS}}$ and the flux through the pentose phosphate pathway (or through the enzyme glucose-6-phosphate dehydrogenase, G6PDH) $\phi_{\text{G6PDH}}\equiv \phi_{\text{PPP}}$. All fluxes can be written in terms of these. In particular, one finds
\begin{equation}
\phi_{{\rm ATPase}}=2u^{{\rm GLC}}-\frac{\phi_{\text{PPP}}}{3}-\phi_{\text{RLS}}~~.
\end{equation}
Now the variation of the {\it extracellular} concentrations of GLC, LAC, K$^+$, Na$^+$ and CO$_2$ due to the operation of a single hRBC is easily seen to be given by
\begin{gather}
\dot{x}^{{\rm GLC}}=-u^{{\rm GLC}}~~~~~,~~~~~
\dot{x}^{{\rm LAC}}=2u^{{\rm GLC}}-\frac{\phi_{\text{PPP}}}{3}\nonumber\\
\dot{x}^{{\rm CO2}}=\phi_{\text{PPP}}~~~~~,~~~~~
\dot{x}^{{\rm Na}}=3\phi_{{\rm ATPase}}\\
\dot{x}^{{\rm K}}=-2\phi_{{\rm ATPase}}~~~.\nonumber
\end{gather}
In turn, for the extracellular medium one has
\begin{equation}
H=\frac{(\dot{x}^{{\rm GLC}})^2}{x^{{\rm GLC}}}+
\frac{(\dot{x}^{{\rm LAC}})^2}{x^{{\rm LAC}}}+
\frac{(\dot{x}^{{\rm CO2}})^2}{x^{{\rm CO2}}}+
\frac{(\dot{x}^{{\rm Na}})^2}{x^{{\rm Na}}}+
\frac{(\dot{x}^{{\rm K}})^2}{x^{{\rm K}}}
\end{equation}
Minimizing this at fixed $u^{{\rm GLC}}$ and $\phi_{\text{RLS}}$ one finds that
\begin{equation}
\phi^{{\rm PPP}}=6(1-a)u^{{\rm GLC}}-3(1-b)\phi_{\text{RLS}}
\end{equation}
where $a$ and $b$ are defined respectively as
\begin{gather}
a=\frac{(x^{{\rm CO2}})^{-1}}{(x^{{\rm CO2}})^{-1}+(x^{{\rm Na}})^{-1}+(4/9)(x^{{\rm K}})^{-1}+(1/9)(x^{{\rm LAC}})^{-1}}\\
b=\frac{(x^{{\rm CO2}})^{-1}+(1/9)(x^{{\rm LAC}})^{-1}}{(x^{{\rm CO2}})^{-1}+(x^{{\rm Na}})^{-1}+(4/9)(x^{{\rm K}})^{-1}+(1/9)(x^{{\rm LAC}})^{-1}}
\end{gather}
One sees that if the concentration of CO$_2$ is much larger than the others (implying $a\simeq 0$) and $\phi_{\text{RLS}}\simeq 0$ then $\phi^{{\rm PPP}}\simeq 6 u^{{\rm GLC}}$ so that the pentose phosphate pathway consumes roughly all of the glucose (the factor 6 is in agreement with the stoichiometry of carbon atoms in GLC and CO$_2$). We can therefore define the fraction of GLC consumption through the PPP by re-scaling $\phi^{{\rm PPP}}$ by $6 u^{{\rm GLC}}$:
\begin{equation}
F=1-a-(1-b)\frac{\phi_{\text{RLS}}}{2 u^{{\rm GLC}}}
\end{equation}
From this it is also immediately clear that, in general conditions for CO$_2$, $F$ varies between $1-a$ (corresponding to $\phi_{\text{RLS}}\simeq 0$) and $b-a$ (corresponding to maximal $\phi_{\text{RLS}}\simeq 2 u^{{\rm GLC}}$). Using the typical concentration values for the above metabolites in the blood, namely $x^{{\rm LAC}}\simeq 10^{-3}$M, $x^{{\rm CO2}}\simeq  2 ~ 10^{-2}$M, $x^{{\rm Na}} \simeq 0.13$M and $x^{{\rm K}} \simeq 5~ 10^{-3}$M one has $a\simeq 0.194$ and $b\simeq 0.625$ so that $0.43\leq F\leq 0.8$. Despite the roughness of the network reconstruction we employed, the bounds just obtained are in remarkable agreement with experimental evidence. On the other hand, using the empirical estimates $ u^{{\rm GLC}}\simeq  3 ~10^{-7}$M/s and $\phi_{\text{RLS}}\simeq 1.4 ~10^{-7}$M/s we find $F\simeq 0.71$.

\section*{Discussion}

In this paper we have derived a variational principle for the NESS of chemical reaction networks, showing that, for time scales over which chemical potentials can be considered constant, stationary non-equilibrium fluxes minimize the function $H$, see (\ref{H}), in which stoichiometry, intakes, outtakes and concentrations appear as parameters. The cost function is closely related to Hopfield models of neural networks. This allows to rephrase such networks as systems of reactions interacting via Hebb-like rules and subject to an external forcing provided by the boundary fluxes. Furthermore $H$ bears two simple physical interpretations. First, minimizing it amounts to finding the flux organizations that keep the overall ``waste'' of chemical species to a minimum, given the  boundary conditions to be satisfied. Second, it equals (modulo a constant) the time derivative of the entropy production per volume: NESS thus correspond to flux configurations such that the entropy production decreases in time at the smallest allowed rate, in line with the ideas exposed in \cite{Kondepudi:1998fk}. We have investigated the implications of such a picture in toy (random) chemical networks -- where the dependence of the solutions on the network's structural parameters can be fully explored, revealing the existence of a phase transition between flux balanced states with $\min H=0$ and states where unbalances emerge -- and in a small real biochemical network, namely the metabolic network of the human red blood cell, a system that is possibly the closest to the theoretical situation described here.

Making contact with stoichiometric models of metabolic networks is relatively straightforward as long as boundary fluxes are taken to be fixed and one does not include additional objective functions that NESS are required to maximize. For instance, the maximization of biomass flux (a frequent optimization criterion for bacterial metabolism \cite{Feist:2010a}) provides a source of entropy production even if one focuses on states with $H=0$ (as in Flux-Balance-Analysis, FBA). On the other hand the theory developed here suggests that the set of local constraints that describes NESS is provided by the minimization of $H$ rather than by taking $H=0$ or $H\geq 0$ a priori. 

An important property that NESS should possess is thermodynamic feasibility, i.e. they should not contain  infeasible loops \cite{palssonloop,palssonvar}. Inspired by the fluctuation theorem \cite{gasp},  we have proposed here a simple dynamical rule (see (\ref{db})) that ensures that the NESS obtained as the minima of H are indeed void of cycles. In general (i.e. when straightforward minimization of $H$ is carried out), it is possible to get rid of infeasible cycles by complementing the variational problem described here with the minimization of the square norm of the flux vector. This is a consequence of the Gordan theorem of alternatives \cite{solodov}: assuming that the matrix $A=(A_i^\mu)$ has full rank, only one of the following systems has a non-trivial solution: (a) $\sum_\mu A_i^\mu g^\mu< 0~~\forall i$ (for $\{g^\mu\}$ real); (b) $\sum_i A_i^\mu k_i=0~~\forall \mu$ (for $\{k_i\geq 0\}$). For a reaction network with stoichiometric coefficients $\xi_i^\mu$ and fluxes $\phi_i$, defining $A_i^\mu=\phi_i\xi_i^\mu$ one sees that system (a) expresses the condition of thermodynamic stability $\phi_i\Delta G_i\leq 0~~\forall i$, whereas system (b) defines thermodynamically infeasible cycles, so that either a flux configuration $\{\phi_i\}$ is thermodynamically feasible or it contains at least one cycle. Now let us consider a steady-state flux configuration satisfying $\sum_i\xi_i^\mu\phi_i\geq u^\mu$ and let us assume it is thermodynamically infeasible, i.e. that system (b) has a solution $\{k_i\}$. We can then construct a new flux configuration $\{\phi_i'\}$ as $\phi_i'=\phi_i+\lambda k_i\phi_i$, with $\lambda$ a constant. Evidently, still $\sum_i\xi_i^\mu\phi_i'\geq u^\mu$. However defining $Q(\{\phi_i\})=\sum_i\phi_i^2$, it is easily seen that choosing $\lambda$ so that $\frac{d}{d\lambda}Q(\{\phi_i'\})=0$ one gets $Q(\{\phi_i'\})< Q(\{\phi_i\})$. In other terms, starting from a thermodynamically infeasible steady-state flux configuration one can construct another steady-state flux configuration whose  total flux is lower. In turn, $Q$ has to be minimum when the flux configuration is thermodynamically feasible. It would be interesting to find a dynamical justification of this criterion. (Notice that flux minimization is an optimality principle frequently used for metabolic network modeling, see e.g. \cite{holzvar}.)

Perhaps not too surprisingly, the formal aspects of this theory present many common traits with those developed over the past decade for the analysis of large games with heterogeneous interacting agents, specifically with Minority Games \cite{mg,revmg}. For obvious reasons, such an analogy shouldn't be stretched and we do not elaborate in detail on it here. In short, however, reactions (or, more properly, enzymes) in chemical networks can be thought to be involved in a competition for the use of a set of possibly limited resources (the different substrates). Whether a reaction can operate or not depends on how much substrate is available to it, i.e. on the overall substrate concentration and on how many other enzymes can bind the same substrates. Each reaction disposes of two `strategies' (or ways to access the set of resources), corresponding to the vectors of its input and output metabolites in the forward and reverse direction, respectively. Such strategies are anti-correlated: substrates in the forward direction are products in the reverse, and vice-versa. Rules like (\ref{dynam}) can then be read as `learning' processes through which enzymes try to anticipate at each time step whether the substrates needed for the forward or reverse processes are most likely to be available, in order for it to operate. This parallel provides an elementary quantitative flavor to the idea that enzymes in biochemical reaction networks compete for the substrates. In absence of boundary fluxes ($h_i=0$ for each $i$ in (\ref{ydot})), the situation is completely equivalent to the Minority Game with anti-correlated strategies studied in \cite{galla}. In this case, the dynamics converges to $H=0$ and $\phi_i=0$ for each $i$, i.e. the system asymptotically reaches chemical equilibrium. As it should be, NESS are induced by non-zero boundary fluxes, i.e. by non-zero fields $h_i$. This additional term turns out to be the main difference between standard anti-correlated Minority Games and the systems discussed here. 

Besides a possible theoretical interest in deepening the analogy just described (e.g. by extending the dynamical approaches employed for the analysis of multi-agent systems \cite{cool} to models of chemical reaction networks), it will be interesting to see how well the variational principle (\ref{min}) describes flux states in real biochemical networks. 

\begin{acknowledgments}
It is a pleasure to thank E. Aurell, W. Bialek, E. Marinari, M. Marsili, I. Perez Castillo and D. Segr\`e for stimulating discussions and suggestions. ADM wishes to thank the Kavli Institute of Theoretical Physics China [Project of Knowledge Innovation Program (PKIP) of Chinese Academy of Sciences, Grant No. KJCX2.YW.W10], the Initiative for Theoretical Sciences at the Graduate Center, City University of New York, and the Abdus Salam International Centre for Theoretical Physics for hospitality during the final stages of this work. This work is supported by the Seed Project DREAM of the Italian Institute of Technology (IIT) and by the joint IIT/Sapienza Nanomedicine Lab. The IIT Platform Computation is gratefully acknowledged. GU is supported by the FIRB grant RBFR08EKEV.
\end{acknowledgments}

\end{document}